# Token Economics in Real-Life: Cryptocurrency and Incentives Design for Insolar's Blockchain Network

Henry M. Kim
*Schulich School of Business*
York University
Toronto, Canada
hmkim@yorku.ca

Marek Laskowski
*Schulich School of Business*
York University
Toronto, Canada
marlas@yorku.ca

Michael Zargham
*BlockScience*
Oakland, USA
zargham@block.science

Hjalmar Turesson
*Schulich School of Business*
York University
Toronto, Canada
hturesson@schulich.yorku.ca

Matt Barlin
*BlockScience*
Oakland, USA
barlin@block.science

Danil Kabanov
*Insolar*
Zug, Switzerland
danil.kabanov@insolar.io

*Abstract*— The study of how to set up cryptocurrency incentive mechanisms and to operationalize governance is token economics. Given the $250 billion market cap for cryptocurrencies, there is compelling need to investigate this topic. In this paper, we present facets of the token engineering process for a real-life 80-person Swiss blockchain startup, Insolar. We show how Insolar used systems modeling and simulation combined with cryptocurrency expertise to design a mechanism to incentivize enterprises and individual users to use their new MainNet public blockchain network. The study showed subsidy pools that incentivize application developers to develop on the network does indeed have the desired positive effect on MainNet adoption. For a startup like Insolar whose success hinge upon how well their model incentivizes various stakeholders to participate on their MainNet network versus that of numerous alternatives, this token economics simulation analysis provides invaluable insights.

## I. Introduction

The core premise of blockchain is straightforward: rather than relying upon a trusted intermediary to proprietarily maintain one ledger of transactions between members of a network, allow all members to maintain their own copies of the same ledger and ensure that all copies are synchronized. This obviates the need for the intermediary, who often exploits its information asymmetry advantage to act in extractive, inefficient, or corrupt ways. A key challenge of this decentralized design is incentives. Even when not behaving exploitatively, the intermediary is well-incentivized to provide centralized services in the form of transaction fees (e.g. Visa or Mastercard), fees charged for data analysis (e.g. Google), and finder's fees (e.g. Uber) [1]. When such services are provided in a decentralized way, an alternative scheme for incentives is needed.

For instance, the Bitcoin network design mitigates the "double spend" problem where a sender simultaneously sends money that they cannot fully cover to two different recipients [2]. The network ensures that one recipient is designated to be the rightful recipient and nullifies the money transfer to the other. A third party that is neither the sender nor recipient verifies this designation. That third party must be incentivized to verify accurately and not collude with the sender or the recipient. Roughly every ten minutes, the network selects a "miner" to add the next block to the Bitcoin blockchain, where the block is comprised of all transactions logged since the last block formation that are verified by the miner to not be "double spent." Currently, a miner receives 12.5 bitcoins, or roughly $125,000 each time a new block is added.

We have thus simplistically described Bitcoin's token (or crypto) economic model. The model lays out the mechanism by which the Bitcoin economy is sustained: Senders and recipients are assured that money cannot be transferred from insufficient funds and third-party miners are incentivized by the promise of receiving ("mining for") bitcoins to provide this assurance.

Bitcoin constitutes an example where token economics is "in the protocol." That is, bitcoins – the network's tokens – are automatically transferred by programmatic rules, and in so doing, the network properly incentivizes stakeholders to collectively behave to sustain its operations. Contrastingly, platforms such as Hyperledger Fabric and R3's Corda do not use cryptocurrencies to effect stakeholder behavior. Rather, these platforms are meant for use in private networks and use traditional group policies "outside the protocol." Private network partners are usually familiar with each other and transact on an ongoing basis; they are inclined to adhere to policies that will sustain their relationships in and outside of the private blockchain.

In this paper, we present a real-life study conducted to develop an "in the protocol" token economics model for Insolar, an 80-person blockchain startup headquartered in Switzerland with offices in five countries. Insolar conducted an Initial Coin Offering (ICO) of their cryptocurrency in December 2017 and raised $42 million [3]. As they are about to launch their public, permissionless, blockchain – the Insolar MainNet – they sought to update their model using applied systems dynamics modeling and simulation. The simulation study led to insights, which were incorporated into a multi-stakeholder token economics model that complements their updated business model.

We believe that the opportunity to investigate a novel real-world application such as token economics modeling provides rare and contributory visibility to the academic community. To that end, our paper is presented as follows. In the next section, we provide related literature, and briefly describe how a token economics model is engineered. Then, we further describe Insolar and detail their specific requirements for token economics modeling. We then present an excerpt of the simulation study and results related to mechanisms for a "subsidy pool," an incentive program to subsidize third parties who provide applications for user to execute on the MainNet. Finally, we provide concluding remarks about how the simulation informed Insolar's design of the token economics mechanism and provide general insights about cryptocurrencies and token economics.

## II. BACKGROUND

We put forth that there are two distinct streams of academic literature regarding designing micro-economies of blockchain-based cryptographic tokens. The first is an economist's perspective as seminally presented in [4]; the primary author, Christian Catalini, is the current Chief Economist for the Facebook-sponsored Libra cryptocurrency system. Some works study incentives specifically for cryptocurrency miners [5]–[7]. Others have studied incentives for various stakeholders in token-based platforms like speculative [8],[9] and utilitarian [10] cryptocurrency investors, platform users [11], and distributed application developers [12]. As would be expected, closed-form economics models underlie this perspective. Even those that apply more nuanced dynamical models [13], [14] generally eschew complex multi-stakeholder analysis in favor of closed-form economics models. Other complementary works include a bibliometric survey of token economics [15] and token economics applied to specific industries [16].

A contrasting perspective that focuses on the systematic development of token-based economies [17][18] is sometimes referred to as token engineering. Note that this is the process for token modeling, not the larger software engineering process that leads to the blockchain system itself. The outputs of the token engineering process are models, results, and documentation that are implemented in the blockchain system. A key difference vis-à-vis the economics perspective is that multi-stakeholder models can be produced via token engineering through use of simulation models that handle complexity albeit without elegant closed-form solutions.

In particular, the predominant simulation technique employed for token engineering, in addition to game theoretically based models [19], [20], is Agent-Based Modeling (ABM) [18]. For instance, ABM of network dynamics analytically shows that Bitcoin network's specific token model leads the network to a stable, self-sustaining state [21] and other efforts simulate the rich nuances of the Bitcoin trading market [22], [23]. For extensions to other cryptocurrency and blockchain design, many efforts have produced generalized frameworks and simulation environments. They include LUNES-blockchain [24], BlockSci [25], Blocksim [26], Simblock [27], MIT's BASIC [28], and cadCAD [29]. For non-Bitcoin contexts, we discover that these efforts generally focus on ABM to simulate technical features of blockchain consensus mechanisms and throughput and assume an underlying token economics model. Our work is unique insofar as it uses simulation to design different features of an economic model and provides details of such simulations in an academic publication. We could speculate that crypto startups do not wish to divulge such information or are not interested in publicly presenting such work even if they have undertaken it. Regardless, providing a worked through example as we do in this paper appears then to be a novel contribution.

For our work of designing Insolar's token economic system, we use cadCAD [29], which entails conceptualizing the token design problem as a differential game problem in which state variable values evolve over time according to some differential equations. Since mathematical modeling for a complex scenario – as is described for Insolar in Figure 1 – is intractable, the conceptualization must first be modeled graphically to especially show feedback mechanisms between key components.

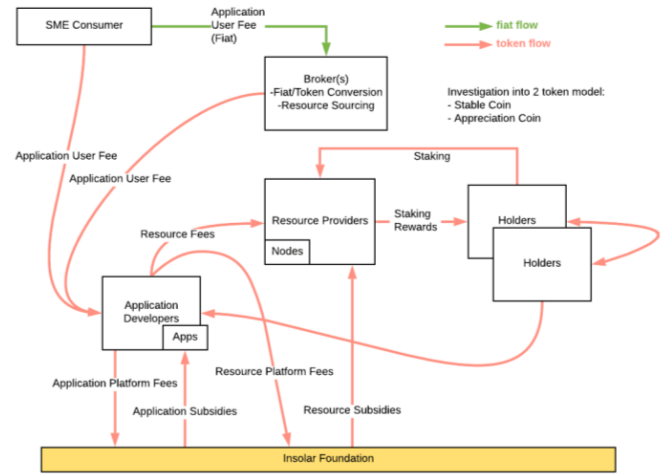

*Figure 1: All feedback mechanisms in Insolar MainNet*

## III. MODELING INSOLAR'S TOKEN ECONOMY

Similar to the Ethereum model, individuals and small enterprises can use the Insolar MainNet as a public, permissionless blockchain for a per transaction fee. Insolar also has a complementary model that does not involve tokens: For a fee in fiat currency, Insolar can design and operate a private blockchain for enterprises in a business network. For instance,

Insolar is involved in other projects in renewable energy, supply chain management, and mining and natural resources.

However, in this paper, our focus is the MainNet. In the following table, we summarize the stakeholders and their roles in the MainNet as modeled in Figure 1 and describe what characteristics the right token economics model would achieve for these roles [30].

TABLE I. ROLES AND MOTIVATIONS WITHIN THE INSOLAR TOKEN ECONOMICS MODEL TYPE STYLES

| Stakeholder | Role in token economic model | The token economics model should: |
|---|---|---|
| SME Consumer (Individual and enterprise users) | • Pays application user fees to Application Developers | • Provide predictable fee structure<br>• Assure predictable quality of application performance |
| Broker(s) | • Converts fiat currency to XNS Coins (MainNet's cryptocurrency) and pays Application Developers on behalf of the SME Consumer | • Ensures "money supply" for MainNet |
| Application Developers | • Develop and deploy Apps (smart contracts) and receive fees for their usage<br>• Pays "rent" for executing applications and using hardware resources to Insolar Foundation as Application Platform and Resource Platform fees, respectively<br>• Also pays Resources Fees to Resource Providers, which in part is used to pay Stake Rewards to XNS holders | • Provide sufficient consumer base and resource capacities<br>• Assure performance quality of Resource Providers consistent with Service Level Agreements (SLA's) |
| Resource Providers | • Provide hardware capacities (telecom, CPU, database) as a Node on the MainNet for running applications and receive fees for this provision<br>• Pay Staking Rewards to XNS Holders | • Provide stable and predictable income for Resource Providers |
| Coin holders | • Provide XNS coins for staking as collateral behind the SLA's and receive Staking Rewards.<br>• Provides stakes to Resource Providers, who in turn extend insurance to Application Developers<br>• | • Provide sufficient staking demand from Resource Providers |
| Insolar Foundation | • Collects Application Platform and Resource Platform Fees and also provides subsidies to Application Developers and Resource Providers to incentivize participation | • Create a large Application Developer community<br>• Create a large hardware resource market<br>• Attract enterprises to use Insolar MainNet<br>• Maintain and improve the platform |

Here are some further clarifications.

- Whereas Bitcoin and Ethereum are peer-to-peer networks that do not pay those that provide CPU cycles, bandwidth, or databases, such Resource Providers are paid in the MainNet. Miners do expend exorbitant amount of resources but that constitutes more of an indirect provision, where contribution is not necessarily proportional to the reward.

- Coin holders provide their XNS's to maintain consensus in the network via staking. Whereas in Bitcoin and Ethereum networks, miners receive tokens to maintain consensus throughout the blockchain (Proof-of-Work), coin holders that have "skin in the game" vote to ensure consensus (Proof-of-Stake). By ensuring that there is a trustworthy consensus mechanism, coin holders/stakers provide confidence to SME Consumers, Application Developers, and Resource Providers, in turn ensuring stability for their coin investments. The coins that are staked serve as a source of insurance for the MainNet: If Application Developers or Resource Providers violate Service Level Agreements (SLA's) with each other, coins from the pool of staked funds can be used to pay the rightfully aggrieved party. Stakers in turn are rewarded a staking fee – an investment income, akin to interest, earned from leaving their staked funds unwithdrawn.

- Not only should the operation of the MainNet be decentralized, so should its governance. A MainNet governed exclusively and in perpetuity by Insolar is not desirable. Instead, a governance foundation, similar to those that underlie the Bitcoin and Ethereum networks and one in which in the long run Insolar is one of many members, is needed.

IV. SIMULATING INSOLAR'S TOKEN MODEL: APPLICATION DEVELOPER SUBSIDY

A. System Simulation Model

The token modeling process starts with a specification of the overall system model, which is shown in Figure 2.

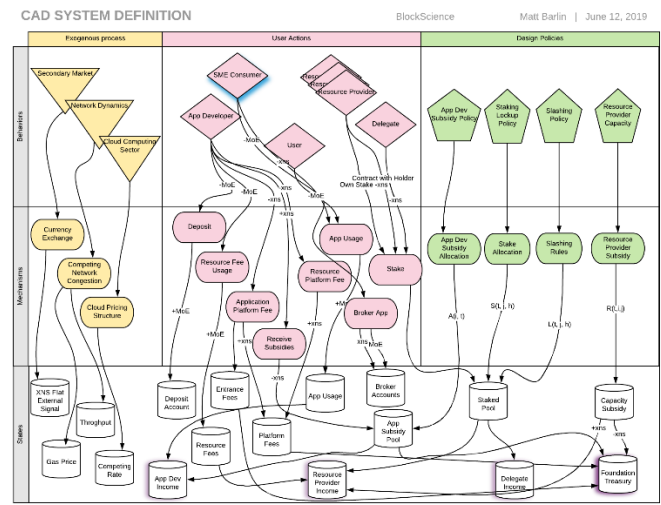

*Figure 2: Overall System Simulation Model for Insolar*

## B. State Transition Model

Although Insolar conducted an extensive study addressing all multiple parties[1], we highlight one facet to bound the scope of this paper: how to incentivize pioneering Application Developers to contribute applications when the MainNet is nascent and network effect has not been achieved yet. The black-outlined excerpt in Figure 1 is a diagram for this excerpt and transitions of states within this excerpt are modeled below in Figure 3.

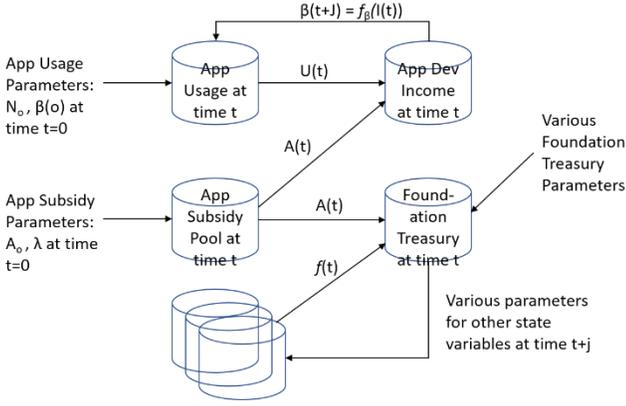

*Figure 3: Transition of State Variables for Insolar (Application Subsidy Excerpt)[2]*

## C. Specification of State Variables

The table below provides a simplified formal specification of the excerpted state variables.

TABLE II. FORMAL SPECIFICATION OF STATE VARIABLES FOR INSOLAR (APPLICATION SUBSIDY EXCERPT)

| Variable | Formal specification |
|---|---|
| App Subsidy Pool | The number of XNS's distributed to developers at time t, $A(t)$, through the App Subsidy is governed by the initial reward pool size $A_0$, and the exponential decay rate $\lambda$. Thus, <br> • $\frac{dA}{dt} = -\lambda \cdot A,$ (1) |
| Number of Apps (App Usage) | Number of apps at time t, $U(t)$, grows at an exponential growth rate $\beta(t)$ from an initial count $N_0$. The growth rate is parametrized to t, $\beta(t)$, whereas the App Subsidy decay rate $\lambda$ is constant. We recognize that number of apps is dependent minimally on numbers of users and developers and other factors. However, we simplify this complex interaction by updating $\beta(t)$ that captures this interaction every time step j as a function solely of app income. This parameter is tuned at the discretion of the modeler, and not necessarily based on formal rules. <br> • $\frac{dU}{dt} = \beta(t) \cdot U,$ (2) <br> $U(t+j)$ is calculated using $\beta(t+j)$, which in turns is modeled as some function $f_\beta$ of the Developer Income $I(t)$ at time t: |
| | • $\beta(t+j) = f_\beta(I(t)).$ (3) |
| App Dev Income | The Application Developer Income (in XNS) at time t, $I(t)$, is the sum of App Subsidy received by time t and the number of applications * c, the average cost of an app. <br> • $I(t) = A_0 - A(t) + c \cdot U(t).$ (4) |
| Foundation Treasury Size | The size of the Foundation Treasury at time t, $T(t)$, is dependent on App Subsidy Pool as well as other state variables such as Capacity Subsidy and Platform Fees that are outside the scope of the subsidy pool excerpt. Moreover, just as how App Dev Income at t is used to update parameters for Number of Apps, $T(t)$ is used to calculate various parameters for variables such as Capacity Subsidy and Platform Fees. $f(t)$ is an arbitrary function of these out-of-the scope variables and $g(t)$ is an arbitrary function of $A(t)$ and $f(t)$. Parameters for other state variables at t+j are used in functions of $T(t)$. <br> • $T(t) = g(A(t), f(t)).$ (5) |

## D. Simulation Runs

App Usage Parameter $N_o$ and $\beta(0)$, and App Subsidy Parameters $A_o$ and $\lambda$ from Table II were the input variables into the simulation. Of these, the key decision variable is $A_o$ as that reflects a policy decision: How much should Insolar initially commit to subsidize Application Developers to develop apps on their platform?

TABLE III. SIMULATION PARAMETERS

| Initial Reward Pool Size in XNS ($A_o$) | Decay Rate[3] ($\lambda$) | Time Steps[4] | Monte Carlo Runs |
|---|---|---|---|
| 250 x 10e6 | 0.0005 | 3652 | 100 |
| 500 x 10e6 | 0.0005 | 3652 | 100 |
| 750 x 10e6 | 0.0005 | 3652 | 100 |
| 1000 x 10e6 | 0.0005 | 3652 | 100 |

The simulations were carried out in a Monte Carlo fashion using four possible scenarios for the starting reward pool size and means of affected state variables – Number of Apps (App Usage), App Dev Income, and Foundation Treasury Size – were computed. The decay rate was kept constant throughout these experiments, as were the numbers of simulated time steps, and Monte Carlo runs.

## E. Simulation Results

Depletion from the Application Subsidy Pool (top panel, Figure 4) confirms the expected exponential decay in token distribution as well as the pattern resulting from the change in initial starting reward pool size. The resulting XNS token distribution to the Application Developers (bottom panel, Figure 4) can be seen in the figure as well. This again is intuitively reasonable given the described model and varying initial reward pool size.

---

[1] MoE in the diagram stands for Medium of Exchange and refers to any arbitrary currency, fiat or crypto, other than XNS tokens

[2] As with Foundation Treasury size, XNS price movements are a function of complex interaction between state variables. As again with Foundation Treasury size, most of the variables that affect XNS prices are outside of the scope of the subsidy excerpt, and so we chose not to discuss simulation of price movements in this paper.

[3] Additional experiments were run during the Insolar economic parameter design and validation research, including but not limited to exploring the effects of varying the decay rate.

[4] Steps in the simulation are in 1 day increments up to 3,652 days or 10 years

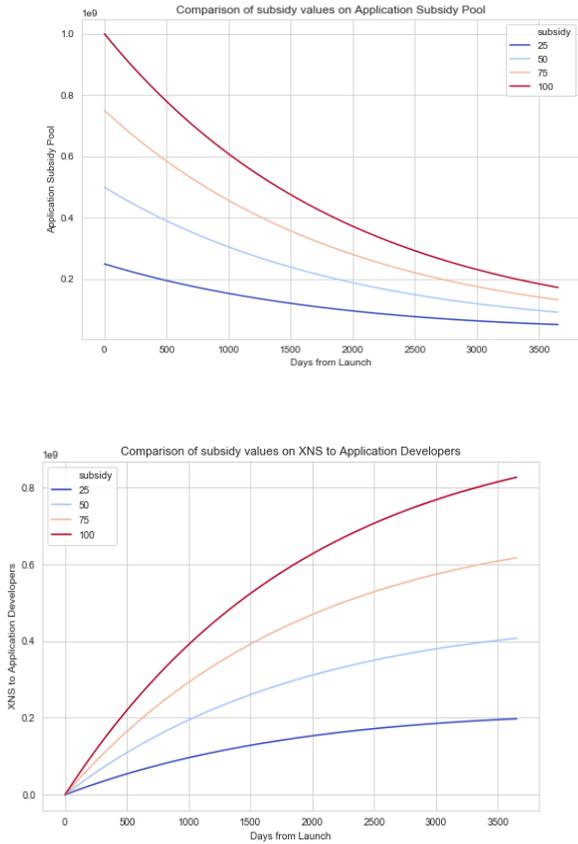

*Figure 4: Subsidy – From Pool and To Developers*

Interestingly, the net effect on Application Developer Income (top panel, Figure 5) is less intuitive. It shows that in the long run Application Developers receive similar incomes regardless of the initial size of the subsidy pool. There appears to be a complex and non-linear relationship between the initial subsidy pool size and reward value delivered to Application Developers.

The initial reward pool size also has an interesting effect on the value of the tokens held by the Insolar Foundation. Predictably, the Foundation's holdings of XNS tokens vary inversely with initial application subsidy pool size. However, due to complex behaviors (from simulation runs related to but outside the scope of the subsidy excerpt), the value of tokens held by the Foundation does increase as initial subsidy pool size is increased (bottom panel, Figure 5).

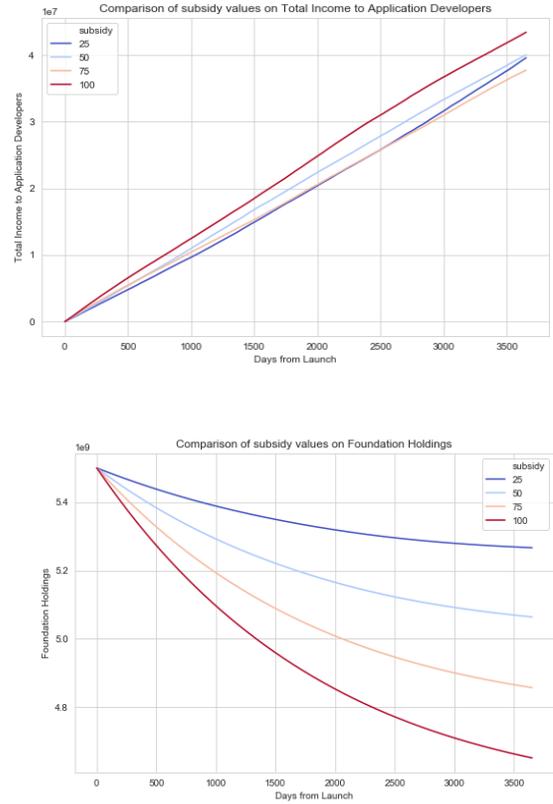

*Figure 5: Subsidy Effect on Application Developer Income and Foundation Treasury*

*F. Simulation Insights*

The key insight from this simulation is that the Application Subsidy has a positive effect on Foundation Treasury; that is, the larger the initial pool of funds for the Application Subsidy, the larger the eventual size of the Foundation Treasury. If we assume that the growth of the treasury is a reasonable by-product of the success of the platform, the simulation results point to the Application Subsidy as having the desired effect of helping platform growth. The results warrant recommending a policy of seeding the Subsidy Pool with significant amount of XNS tokens.

The direct overall benefit to Application Developers is less clear. Even if more XNS tokens are granted to Application Developers through the subsidy, the net effect on them is not guaranteed at every point in time. Application Developers pay Application Platform and Resource Platform fees in XNS tokens and these grow at some rate, yet the rate at which they receive subsidies decays. Ultimately, the Insolar Foundation reaps the reward from this mismatch of accelerating fee revenues and decelerating subsidy outlays.

V. CONCLUDING REMARKS

To summarize, given the increasing role of cryptocurrencies – especially in light of recent announcements of corporate cryptocurrencies from Facebook, JP Morgan, and Walmart [31] – we determine that the topic of token economics is under-researched. This paper presents excerpts of the token engineering process to demonstrate how simulation modelling informed policy design for Insolar's subsidy pool – their mechanism to incentivize application providers to develop more applications onto the public MainNet. Simulation results point to investment into the subsidy pool as having a desirable positive effect that helps MainNet growth. The insights from modeling will be important as Insolar tries to foster a vibrant ecosystem of application developers and computing resource providers that will entice individual and enterprise users to join the MainNet after launch.

For this brief paper, a fuller exploration of theoretical issues was beyond our scope. We did not delve much into the formal control theory underlying these simulation models, nor did we relate much to traditional literature on market design from Economics and Operations Research. Moreover, though always a pressing issue for blockchain-based systems, we did not address security issues as bounding of our scope entailed assuming security breaches to the MainNet to be exogenous events that our simulation does not factor in. Regardless of these limitations, for a fledgling startup like Insolar, their future success will in no small part hinge upon how well their model incentivizes various stakeholders to participate on their MainNet network versus that of numerous alternatives. We believe that providing visibility to such a strategic and academically novel exercise serves a contribution to the research community.

Practically, Insolar is also interested in future work that more granularly examines application subsidies, namely the mix of spending between these subsidy types:

- *App Seeding* entails incentivizing application creation events such as "code jams" or "hackathons."
- *New Application Subsidy* is offered for newly deployed applications, with the subsidy decreasing to 0 the developer after an "introductory discount" time.
- A *Volume Resource Subsidy* is meant to ease the resources costs incurred by developers of popular applications.
- A *Success Reward* is awarded to Application Developers for meeting strategic Key Performance Indicators, as determined by the Insolar Foundation.

The simulation model required for determining this mix needs to build upon and extend the work presented in this paper.

## VII. AUTHOR BIOS

Henry Kim is an Associate Professor at the Schulich School of Business and the Director of the blockchain.lab at York University. Prof. Kim has written extensively on blockchain, authoring more than 25 blockchain-related papers in venues such as IEEE journals and Frontiers in Blockchain. He serves as a senior research advisor for several blockchain startups and co-organized the 2$^{nd}$ IEEE Conference on Blockchain and Cryptocurrencies. Prof. Kim received his PhD from the University of Toronto. He is a member of the IEEE.

Marek Laskowski serves as Senior Research Scientist with BlockScience and is a Co-Founder of blockchain.lab and Toronto Blockchain Week. He also advises several blockchain startups as well as UN/CEFACT. He holds a Ph.D. in Computer Engineering from the University of Manitoba, where his work on multi-agent modelling of complex healthcare systems was cited by the US Centers for Disease Control (CDC).

Michael Zargham is the founder and CEO of BlockScience, an engineering, R&D, and analytics firm combining systems theory, computational social science to design economic systems. He holds a PhD in Systems Engineering from the University of Pennsylvania where he worked on dynamic resource allocation in decentralized networks.

Hjalmar Turesson was born in Helsingborg, Sweden. He received a B.Sc. in Biology from Lund University, Sweden, a M.Sc. from the University of Tübingen, Germany, and a Ph.D. from Princeton University. Currently, he is teaching AI at Schulich School of Business, York University, Canada.

Matthew Barlin is Lead Systems Engineer at BlockScience with research interests in Model Based Systems Engineering and blockchain. He holds an SM from MIT in Ocean Systems Management and is a member of INFORMS.

Danil Kabanov is the Head of Development Center at Insolar. His research interest are in Enterprise Blockchain Systems. He received a Master of Mathematics (Crypto analysis) from Tomsk State University and a Master of Economics from St. Petersburg Polytechnic University.